\documentclass[conference]{IEEEtran}
\IEEEoverridecommandlockouts

\usepackage{graphicx}
\usepackage{pgf}
\usepackage{tikz}
\usetikzlibrary{arrows}
\usetikzlibrary{arrows,automata}
\usetikzlibrary{spy}
\usetikzlibrary{patterns}
\usepackage[latin1]{inputenc}
\usepackage{subcaption}
\usepackage{amssymb}
\usepackage{tabularx,booktabs}
\usepackage{pgfplots}
\usepackage{lipsum}
\usepackage{dsfont}
\usepackage{tikz-layers}
\usepackage{setspace}
\usepackage{svg}
\usepackage{subcaption}
\usepackage[ruled,vlined]{algorithm2e}
\pgfplotsset{compat=1.6}
\usepgfplotslibrary{statistics}
\usepgfplotslibrary{colorbrewer} 
\pgfplotsset{compat = 1.15, cycle list/Set1-8} 
\usetikzlibrary{pgfplots.statistics, pgfplots.colorbrewer} 
\usepackage{pgfplotstable}
\usepackage{filecontents}
\usepackage{amsmath}
\usepackage{bm}
\usepackage{mathtools}
\usepackage[hyphens]{url}
\usepackage[hidelinks]{hyperref}

\usepackage{comment}
\usepackage[acronym,toc]{glossaries}
\usepackage{glossaries-extra}
\setabbreviationstyle[acronym]{long-short}
\GlsXtrEnableEntryCounting
 {acronym}
 {1}
\makeglossaries

\usepackage{enumitem}
\usepackage{bbm}
\usepackage[cmintegrals]{newtxmath}
\usepackage{stackengine}
\usepackage{afterpage}
\usepackage[euler]{textgreek}
\usepackage{lgreek}
\usepackage{cite}

\def\BibTeX{{\rm B\kern-.05em{\sc i\kern-.025em b}\kern-.08em
    T\kern-.1667em\lower.7ex\hbox{E}\kern-.125emX}}

\usetikzlibrary{plotmarks}
\usetikzlibrary{arrows.meta}
\usepackage{pgfplots}

\newcommand{\calB}{\mathcal{B}}

\newcommand{\calU}{\mathcal{U}}

\definecolor{mycolor1}{rgb}{0.29, 0.59, 0.82}%
\definecolor{mycolor2}{rgb}{1.0, 0.4, 0.6}
\definecolor{mycolor3}{rgb}{0.92900,0.69400,0.12500}%
\definecolor{mycolor4}{rgb}{0.71,0.49,0.86}
\definecolor{mycolor5}{rgb}{0.12, 0.3, 0.17}
\definecolor{mycolor6}{rgb}{0.43, 0.21, 0.1}
\definecolor{mycolor7}{rgb}{0.52, 0.73, 0.4}
\definecolor{mycolor8}{rgb}{0.98, 0.38, 0.5}
\definecolor{vColor}{rgb}{0.12, 0.3, 0.17}

\newacronym{TV}{TV}{television}
\newacronym{DTTB}{DTTB}{digital television terrestrial broadcasting}
\newacronym{DVB}{DVB}{Digital Video Broadcast}
\newacronym{DVB-H}{DVB-H}{Digital Video Broadcast-Handheld}
\newacronym{ATSC}{ATSC}{Advanced Television System Committee}
\newacronym{ATSC-M/H}{ATSC-M/H}{Advanced Television System Committee - Mobile/Handheld}
\newacronym{IPTV}{IPTV}{Internet Protocol television}
\newacronym{IP}{IP}{Internet Protocol}

\newacronym{UE}{UE}{user equipment}
\newacronym{PC}{PC}{personal computer}
\newacronym{RAN}{RAN}{radio access network}
\newacronym{CN}{CN}{core network}
\newacronym{MS}{MS}{mobile station}

\newacronym{ITU-R}{ITU-R}{International Telecommunications Union - Radiocommunication Sector}
\newacronym{IMT-Advanced}{IMT-Advanced}{International Mobile Telecommunications Advanced}
\newacronym{4G}{4G}{fourth-generation of mobile phone communications and Internet access technology}
\newacronym{3gpp}{3GPP}{3rd Generation Partnership Project}

\newacronym{GSM}{GSM}{Global System for Mobile Communications}
\newacronym{UMTS}{UMTS}{Universal Mobile Telecommunications System}
\newacronym{HSPA}{HSPA}{High Speed Packet Access}
\newacronym{lte}{LTE}{Long-Term Evolution}
\newacronym{lte-a}{LTE-A}{Long-Term Evolution Advanced}

\newacronym{e-UTRAN}{e-UTRAN}{evolved Universal Terrestrial Radio Access Network}
\newacronym{eNB}{eNB}{e-UTRAN NodeB}
\newacronym{gNB}{gNB}{gNodeB}
\newacronym{EPC}{EPC}{Evolved Packet Core}

\newacronym{MBMS}{MBMS}{Multimedia and Broadcast Multicast Service}
\newacronym{eMBMS}{eMBMS}{Evolved MBMS}
\newacronym{SFN}{SFN}{single-frequency network}
\newacronym{MBSFN}{MBSFN}{MBMS single-frequency network}
\newacronym{BM-SC}{BM-SC}{Broadcast/Multicast Service Center}
\newacronym{MBMS GW}{MBMS GW}{MBMS Gateway}
\newacronym{MME}{MME}{Mobility Management Entity}
\newacronym{MCE}{MCE}{Multi-cell/multicast Coordinating Entity}
\newacronym{SYNC}{SYNC}{synchronization}
\newacronym{MCCH}{MCCH}{Multicast Control Channel}
\newacronym{MTCH}{MTCH}{Multicast Traffic Channel}
\newacronym{MCH}{MCH}{Multicast Channel}
\newacronym{PMCH}{PMCH}{Physical Multicast Channel}
\newacronym{PDSCH}{PDSCH}{Physical Downlink Shared Channel}

\newacronym{IEEE}{IEEE}{Institute of Electrical and Electronics Engineers}
\newacronym{WiMAX}{WiMAX}{Worldwide Interoperability for Microwave Access}
\newacronym{ASN}{ASN}{access service network}
\newacronym{ASN-GW}{ASN-GW}{ASN gateway}
\newacronym{CSN}{CSN}{Connectivity Service Network}

\newacronym{PA}{PA}{power amplifier}

\newacronym{NI}{NI}{National Instruments}

\newacronym{TDD}{TDD}{time-division duplex}
\newacronym{FDD}{FDD}{frequency-division duplex}
\newacronym{UDP}{UDP}{User Datagram Protocol}
\newacronym{APP}{APP}{application}
\newacronym{mac}{MAC}{medium access control}
\newacronym{phy}{PHY}{physical}
\newacronym{RLC}{RLC}{radio link control}
\newacronym{FIFO}{FIFO}{first-in first-out}
\newacronym{CRC}{CRC}{cyclic redundancy check}
\newacronym{SAP}{SAP}{service access point}
\newacronym{FEC}{FEC}{forward error correction}
\newacronym{IF}{IF}{intermediate frequency}
\newacronym{RF}{RF}{radio frequency}
\newacronym{mimo}{MIMO}{multiple-input and multiple-output}

\newacronym{SPC}{SPC}{superposition coding}
\newacronym{SVC}{SVC}{Scalable Video Coding}
\newacronym{GM}{GM}{generic multicasting}
\newacronym{SCM}{SCM}{superposition coded multicasting}
\newacronym{SIC}{SIC}{successive interference cancellation}

\newacronym{st}{ST}{secondary transmitter}
\newacronym{pt}{PT}{primary transmitter}
\newacronym{sr}{SR}{secondary receiver}
\newacronym{pr}{PR}{primary receiver}
\newacronym{su}{SU}{secondary user}
\newacronym{pu}{PU}{primary user}

\newacronym{awgn}{AWGN}{additive white Gaussian noise}

\newacronym{pdf}{PDF}{probability density function}
\newacronym{iid}{i.i.d.}{independent and identicaly distributed}
\newacronym{rf}{RF}{radio frequency}

\newacronym{dd}{DD}{Device-to-Device}
\newacronym{ddu}{DDU}{Device-to-Device user}
\newacronym{dds}{DDS}{Device-to-Device system}
\newacronym{ddt}{DT}{DDU transmitter}
\newacronym{ddr}{DR}{DDU receiver}

\newacronym{bs}{BS}{base station}
\newacronym{bsu}{BSU}{base station associated user}
\newacronym{bss}{BSS}{base station associated system}
\newacronym{bst}{BT}{BSU transmitter}
\newacronym{bsr}{BR}{BSU receiver}

\newacronym{epg}{EPG}{energy per goodbit}
\newacronym{mepg}{MEPG}{modified energy per goodbit}
\newacronym{ee}{EE}{energy efficiency}
\newacronym{se}{SE}{spectral efficiency}

\newacronym{wrt}{w.r.t.}{with respect to}

\newacronym{kkt}{KKT}{Karush-Kuhn-Tucker}
\newacronym{admm}{ADM}{Alternating Directing Method}
\newacronym{cr}{CR}{cognitive radio}
\newacronym{ssi}{SSI}{soft-sensing information}
\newacronym{csi}{CSI}{channel state information}
\newacronym{qsi}{QSI}{queue state information}
\newacronym{el}{EL}{enhancement layer(s)}
\newacronym{snr}{SNR}{signal-to-noise ratio}

\newacronym{NAL}{NAL}{network abstraction layer}
\newacronym{QP}{QP}{quantization parameter}

\newacronym{ofdm}{OFDM}{orthogonal frequency-division multiplexing}
\newacronym{ofdma}{OFDMA}{orthogonal frequency-division multiple access}
\newacronym{tdma}{TDMA}{time division multiple access}

\newacronym{PUSC}{PUSC}{partial usage of the subchannels}
\newacronym{CFO}{CFO}{carrier frequency offset}
\newacronym{I/Q}{I/Q}{in-phase and quadrature-phase}
\newacronym{ASK}{ASK}{amplitude-shift keying}
\newacronym{PSK}{PSK}{phase-shift keying}
\newacronym{BPSK}{BPSK}{binary phase-shift keying}
\newacronym{QPSK}{QPSK}{quadrature phase-shift keying}
\newacronym{QAM}{QAM}{quadrature amplitude modulation}
\newacronym{PSNR}{PSNR}{peak signal-to-noise ratio}
\newacronym{PELR}{PELR}{packet error and loss rate}

\newacronym{kNN}{\textit{k}-NN}{\textit{k}-nearest neighbor algorithm}
\newacronym{SVM}{SVM}{support vector machines}
\newacronym{nn}{NN}{neural network}
\newacronym{NN}{NN}{neural network}
\newacronym{dnn}{DNN}{deep neural network}
\newacronym{RBF}{RBF}{radial basis function}
\newacronym{RMSE}{RMSE}{root mean squared error}
\newacronym{mse}{MSE}{mean squared error}
\newacronym{lmse}{LMSE}{linear mean square-error estimator}

\newacronym{R2}{$R^2$}{coefficient of determination}

\newacronym{KAUST}{KAUST}{King Abdullah University of Science and Technology}
\newacronym{GSA}{GSA}{Global mobile Suppliers Association}

\newacronym{VoD}{VoD}{video on demand}
\newacronym{HEVC}{HEVC}{High Efficiency of Video Coding}
\newacronym{DASH}{DASH}{Dynamic Adaptive Streaming over HTTP}

\newacronym{PUT}{PUT}{people using television}

\newacronym{ADTVS}{ADTVS}{Audience Driven live TV Scheduling}

\newacronym{arq}{ARQ}{automatic repeat request}

\newacronym{harq}{HARQ}{hybrid automatic repeat request}

\newacronym{sdp}{SDP}{semi-definite programming}

\newacronym{tcp}{TCP}{transmission control protocol}

\newacronym{e2e}{E2E}{end-to-end}

\newacronym{ran}{RAN}{radio access network}
\newacronym{cran}{CRAN}{cloud radio access network}
\newacronym{udcran}{UD-CRAN}{ultra-dense CRAN}
\newacronym{dran}{DRAN}{distributed radio access network}
\newacronym{hcran}{H-CRAN}{hybrid cloud radio access network}
\newacronym{hetnet}{HetNet}{heterogeneous network}
\newacronym{vcran}{V-CRAN}{virtualized CRAN}
\newacronym{ecran}{E-CRAN}{edge-CRAN}
\newacronym{hvcran}{H-VCRAN}{hybrid-virtualized CRAN}

\newacronym{bbu}{BBU}{baseband processing unit}
\newacronym{rrh}{RRH}{remote radio head}
\newacronym{ru}{RU}{radio unit}
\newacronym{rs}{RS}{remote site}
\newacronym{cs}{CS}{central site}

\newacronym{rru}{RRU}{radio resource unit}
\newacronym{rb}{RB}{resource block}
\newacronym{hpn}{HPN}{high-power node}
\newacronym{lpn}{LPN}{low-power node}
\newacronym{mabs}{MaBS}{macro basestation}
\newacronym{ue}{UE}{user equipment}

\newacronym{comp}{CoMP}{coordinated multi-point}
\newacronym{ranaas}{RANaaS}{RAN-as-a-Service}

\newacronym{rof}{RoF}{radio over fiber}
\newacronym{wdm}{WDM}{Wavelength Division Multiplexing}
\newacronym{dls}{DLS}{distributed large scale}

\newacronym{qos}{QoS}{quality of service}
\newacronym{qoe}{QoE}{quality of experience}
\newacronym{qee}{QEE}{quality of energy-efficiency}

\newacronym{gg}{GG}{group-to-group}
\newacronym{ht}{HT}{hyper-transceiver}

\newacronym{fh}{FH}{fronthaul}
\newacronym{dl}{DL}{downlink}
\newacronym{ul}{UL}{uplink}

\newacronym{cp}{CP}{Cell-Processing}
\newacronym{up}{UP}{User-Processing}

\newacronym{co}{CO}{center office}

\newacronym{du}{DU}{digital unit}
\newacronym{lc}{LC}{Line-Card}

\newacronym{onu}{ONU}{optical network unit}
\newacronym{olt}{OLT}{optical line terminal}
\newacronym{osw}{OSW}{optical switch}

\newacronym{es}{ES}{ethernet switch}

\newacronym{ppp}{PPP}{Poisson point process}

\newacronym{mppp}{MPPP}{marked Poisson point process}

\newacronym{sinr}{SINR}{signal to noise and interference ratio}

\newacronym{sir}{SIR}{signal to interference ratio}

\newacronym{mbs}{MBS}{macro basestation}
\newacronym{ap}{AP}{access point}
\newacronym{fap}{FAP}{femto-cell access point}
\newacronym{sap}{SAP}{small-cell access point}
\newacronym{iot}{IoT}{Internet of Things}
\newacronym{ti}{TI}{Tactile Internet}
\newacronym{lsm}{LSM}{linear scalarizing method}

\newacronym{lp}{LP}{Low-Priority}
\newacronym{hp}{HP}{High-Priority}
\newacronym{lpu}{LPU}{Low-Priority user}
\newacronym{hpu}{HPU}{High-Priority user}
\newacronym{lps}{LPS}{Low-Priority system}
\newacronym{hps}{HPS}{High-Priority system}

\newacronym{ttm}{TTM}{time to market}
\newacronym{udn}{UDN}{ultra-dense network}

\newacronym{capex}{CAPEX}{capital expenditure}
\newacronym{opex}{OPEX}{operational expenditure}

\newacronym{cpri}{CPRI}{common public radio interface}
\newacronym{otn}{OTN}{optical transport network}
\newacronym{pon}{PON}{passive optical network}
\newacronym{twdm}{TWDM}{time and wavelength division multiplexing}

\newacronym{ec}{EC}{Edge-Cloud}
\newacronym{cc}{CC}{Central-Cloud}

\newacronym{mmw}{m-Wave}{Milli-Meter wave}

\newacronym{gops}{GOPS}{giga operation per second}
\newacronym{mops}{MOPS}{mega operation per second}

\newacronym{ip}{IP}{internet protocol}
\newacronym{rlc}{RLC}{radio link control}
\newacronym{pdcp}{PDCP}{packet data convergence protocol}

\newacronym{mno}{MNO}{mobile network operator}
\newacronym{prb}{PRB}{physical resource block}
\newacronym{mi}{MI}{modulation index}

\newacronym{wifi}{WiFi}{wireless local area network}
\newacronym{cpu}{CPU}{central processing unit}
\newacronym{vcpu}{VCPU}{virtual CPU}
\newacronym{vm}{VM}{virtual machine}

\newacronym{urs}{UrS}{user requested service}

\newacronym{rsf}{RSF}{radio sub-frame}
\newacronym{siso}{SISO}{single-input single-output}

\newacronym{mec}{MEC}{mobile edge computing}
\newacronym{co2}{CO$_{2}$}{carbo dioxide}

\newacronym{ar}{AR}{augmented reality}
\newacronym{vr}{VR}{virtual reality}
\newacronym{cfp}{CFP}{communication function processing}
\newacronym{ptp}{PTP}{precision time protocol}

\newacronym{voip}{VoIP}{voice over Internet protocol}

\newacronym{da}{DA}{data analytics}

\newacronym{kpi}{KPI}{key performance indicator}

\newacronym{fsmc}{FSMC}{finite state markov chain}
\newacronym{ml}{ML}{machine learning}

\newacronym{5g}{5G}{fifth generation of mobile communication systems}
\newacronym{gnbcu}{gNB-CU}{gNB central unit}
\newacronym{gnbdu}{gNB-DU}{gNB distributed unit}
\newacronym{ecpri}{eCPRI}{common public radio interface}
\newacronym{fl}{FL}{federated learning}
\newacronym{rsrq}{RSRQ}{reference signal received quality}
\newacronym{rsrp}{RSRP}{reference signal received power}
\newacronym{urllc}{URLLC}{ultra-reliable low-latency communications}
\newacronym{embb}{eMBB}{enhanced mobile broadband}

\newacronym{mae}{MAE}{modified autoencoder}
\newacronym{mtc}{MTC}{machine type communication}
\newacronym{mmtc}{mMTC}{massive machine type communication}
\newacronym{pca}{PCA}{principal component analysis}

\newacronym{cps}{CPS}{cyber-physical system}
\newacronym{gnb}{gNB}{gNodeB}
\newacronym{ref}{REF}{reliability enhancement feature}
\newacronym{nfo}{NFO}{network level feature orchestrator}
\newacronym{dc}{DC}{data center}
\newacronym{nf}{NF}{network function}
\newacronym{vnf}{VNF}{virtual network function}
\newacronym{nfv}{NFV}{network functions virtualization}
\newacronym{nssmf}{NSSMF}{network slice subnet management function}
\newacronym{nsmf}{NSMF}{network slice management function}
\newacronym{ai}{AI}{artificial intelligence}
\newacronym{rl}{RL}{reinforcement learning}
\newacronym{hrl}{HRL}{hierarchical reinforcement learning}
\newacronym{ddpg}{DDPG}{deep deterministic policy gradient}
\newacronym{dqn}{DQN}{deep Q-networks}
\newacronym{sac}{SAC}{soft actor-critic}
\newacronym{a2c}{A2C}{advantage actor-critic}
\newacronym{bsac}{BSAC}{branching SAC}
\newacronym{td3}{TD3}{twin delayed deep deterministic policy gradient algorithm}
\newacronym{sgd}{SGD}{stochastic gradient descent}
\newacronym{um}{UM}{unacknowledged mode}
\newacronym{am}{AM}{acknowledged mode}

\newacronym{tsn}{TSN}{time-sensitive networks}
\newacronym{nr}{NR}{New radio}

\newacronym{inf}{InF}{indoor factory}

\newacronym{ttl}{TTI}{transmission time interval}

\newacronym{pdcch}{PDCCH}{physical downlink control channel}
\newacronym{dmrs}{DMRS}{demodulation reference signal}
\newacronym{pi}{PI}{puncturing indication}
\newacronym{scs}{SCS}{subcarrier spacing}

\newacronym{gf}{GF}{grant-free}
\newacronym{rrc}{RRC}{radio resource control}
\newacronym{sps}{SPS}{semi-persistent scheduling}

\newacronym{ack}{ACK}{acknowledgement}
\newacronym{nack}{NACK}{negative-acknowledgement}

\newacronym{cqi}{CQI}{channel quality indicator}
\newacronym{mcs}{MCS}{modulation and coding scheme}

\newacronym{mdp}{MDP}{Markov decision process}

\newacronym{td}{TD}{temporal difference}

\newacronym{a3c}{A3C}{asynchronous advantage actor-Critic}

\newacronym{msbe}{MSBE}{mean-squared Bellman error}

\newacronym{smdp}{SMDP}{semi-Markov decision process}

\newacronym{lphrl}{LP-HRL}{learning hierarchical policy HRL}
\newacronym{unihrl}{UNI-HRL}{learning hierarchical policy in unification with subtask discovery HRL}
\newacronym{thrl}{T-HRL}{transfer learning with HRL}
\newacronym{isdhrl}{ISD-HRL}{independent subtask discovery HRL}
\newacronym{mahrl}{MA-HRL}{multi-agent HRL}

\newacronym{hiro}{HIRO}{hierarchical reinforcement learning with off-policy correction}

\newacronym{hams}{HAMs}{hierarchy of abstract machines}
\newacronym{fnn}{FnN}{feudal networks}
\newacronym{oc}{OC}{option critic}
\newacronym{dac}{DAC}{double actor critic}

\newacronym{marl}{MA-RL}{multi-agent reinforcement learning}

\newacronym{isemo}{ISEMO}{inter subtask empowerment based multi-agent options}
\newacronym{iser}{ISER}{inter subtask empowerment rewards}

\newacronym{fmh}{FMH}{feudal multi-agent hierarchies}

\newacronym{doc}{DOC}{distributed options critic}

\newacronym{hdqn}{h-DQN}{hierarchical deep Q-network}
\newacronym{uav}{UAV}{unmanned aerial vehicle}
\newacronym{hddpg}{H-DDPG}{hierarchical-DDPG}

\newacronym{ht3o}{HT3O}{hierarchical trajectory optimizatffion and offloading optimization}

\newacronym{mhrl}{MHRL}{meta-hierarchical reinforcement learning}

\newabbreviation[description={\glsxtrshort{inf}-with dense clutter and high base station height}]{inf-dh}{InF-DH}{indoor factory with dense clutter and high base station height}

\newabbreviation{cdf}{CDF}{cumulative distribution function}
\newabbreviation{ccdf}{CCDF}{complementary CDF}

\begin{document}
\begin{NoHyper}
\bstctlcite{IEEEexample:BSTcontrol}	
\title{Communication-Efficient Orchestrations for URLLC Service via Hierarchical Reinforcement Learning} 

\author{
\IEEEauthorblockN{Wei Shi\IEEEauthorrefmark{1}\IEEEauthorrefmark{2}, Milad Ganjalizadeh\IEEEauthorrefmark{1}\IEEEauthorrefmark{2}, Hossein Shokri Ghadikolaei\IEEEauthorrefmark{1}, and Marina Petrova\IEEEauthorrefmark{2}\IEEEauthorrefmark{3}
\thanks{This work was partly supported by Swedish Foundation for Strategic Research (SSF) under Grant iPhD:ID17-0079.}
\thanks{\copyright 2023 IEEE. Personal use of this material is permitted. Permission from IEEE must be obtained for all other uses, in any current or future media, including reprinting/republishing this material for advertising or promotional purposes, creating new collective works, for resale or redistribution to servers or lists, or reuse of any copyrighted component of this work in other works.}
}
%
\vspace{3mm}
\IEEEauthorblockA{
\IEEEauthorrefmark{1}Ericsson Research, Sweden\\
\IEEEauthorrefmark{2}School of Electrical Engineering and Computer Science, KTH Royal Institute of Technology, Stockholm, Sweden\\
\IEEEauthorrefmark{3}Mobile Communications and Computing, RWTH Aachen University, Germany
\\
Email:
\{wei.b.shi, milad.ganjalizadeh, hossein.shokri.ghadikolaei\}@ericsson.com,
petrovam@kth.se}
\vspace{-1cm}
}
\markboth{}%
{Shell \MakeLowercase{\textit{et al.}}: Bare Demo of IEEEtran.cls for IEEE Communications Society Journals}

\maketitle
	
\begin{abstract}
Ultra-reliable low latency communications (URLLC) service is envisioned to enable use cases with strict reliability and latency requirements in 5G. 
One approach for enabling URLLC services is to leverage Reinforcement Learning (RL) to efficiently allocate wireless resources.
However, with conventional RL methods, the decision variables (though being deployed at various network layers) are typically optimized in the same control loop, leading to significant practical limitations on the control loop's delay as well as excessive signaling and energy consumption.
In this paper, we propose a multi-agent Hierarchical RL (HRL) framework that enables the implementation of multi-level policies with different control loop timescales. Agents with faster control loops are deployed closer to the base station, while the ones with slower control loops are at the edge or closer to the core network providing high-level guidelines for low-level actions. On a use case from the prior art, with our HRL framework, we optimized the maximum number of retransmissions and transmission power of industrial devices.
Our extensive simulation results on the factory automation scenario show that the HRL framework achieves better performance as the baseline single-agent RL method, with significantly less overhead of signal transmissions and delay compared to the one-agent RL methods.
\end{abstract}
	
\IEEEpeerreviewmaketitle

\section{Introduction}\label{sec:intro}
Nowadays, the development of \gls{5g} technology has achieved a mature technical standard aiming to provide wireless communication services to multiple vertical industrial areas \cite{ETRservice}. According to \gls{3gpp}~\cite{3gpp.22.261}, \gls{urllc} stands as one of the three main services for \gls{5g} standards, and beyond the standardization, it has shown significant improvements in the efficiency and performance of communication systems \cite{li20185g, popovski2019wireless, ren2020joint}. 
The main requirements for \gls{urllc} (especially in the context of \glspl{cps}) are high reliability (e.g., 10 years without failure), high availability (e.g., 99.9999\%), and low latency (below some tens of milliseconds). \Gls{ml} has proven effective in meeting these stringent requirements over resource-limited and faulty wireless channels~\cite{azari2019risk,kasgari2019model,lillicrap2015continuous,ganjalizadeh2021rl}.

\subsection{Literature Review}\label{se:1.1.1}
Various \gls{ml}-based optimization schemes have been proposed for \gls{urllc}. For example, reference \cite{azari2019risk} proposes a distributed risk-sensitive \gls{ml} solution for hybrid orthogonal/non-orthogonal radio resource slicing, regulating the spectrum to satisfy the \gls{urllc} requirements. 
Reference \cite{kasgari2019model} implements an \gls{rl} framework with the \gls{ddpg} algorithm \cite{lillicrap2015continuous} into an \gls{ofdma} system, minimizing the transmission power. Reference \cite{ganjalizadeh2021rl} optimizes both power and \gls{harq} retransmission scheme, leading to further improvements in terms of reliability and availability in factory automation use cases. 
However, adopting these strategies in real-life applications could be impractical:

\begin{itemize}[leftmargin=*]
    \item  Conventional (flat) RL methods only have one action vector that forces all decision variables to be designed at the same control loop (e.g., resource blocks and power allocation in \cite{kasgari2019model} and power and \gls{harq} retransmission scheme in \cite{ganjalizadeh2021rl}). However, \gls{5g} services usually require various decision variables to be tuned on different control loops. For example, we prefer to have one fixed slicing decision for many coherence intervals while we can constantly adjust the transmit power at every coherence interval. 
    \item To make a decision in a flat RL, we need to collect the information required to determine the state including the ones corresponding to variables with a much slower control loop (slicing in our example). The interval of such data collection is determined by the faster control loop. Such unnecessary data collection results in more energy consumption, network congestion, and extra latency, which could be detrimental to the sustainable operation of \gls{urllc} service. 
\end{itemize}

\noindent To address these limitations, we develop a \gls{hrl} framework to optimize the operation of URLLC. 

\subsection{Hierarchical Reinforcement Learning}\label{se:1.1.3}

Different from one-level \gls{rl} methods, \gls{hrl} decomposes one \gls{rl} problem into a hierarchy of sub-problems, which allows optimizing different tasks independently with different algorithms, timescales, models, and multiple agents \cite{pateria2021hierarchical}. This usually brings a reduction in exploration complexity, computation, signaling, and time required for the training and inference processes. Reference \cite{shah2021joint} proposes a hierarchical deep actor-critic method for the resource allocation problems of the 6G massive \gls{iot} scenarios. Reference \cite{liu2021dynamic} introduces a hierarchical \gls{dqn} model with one main controller and multiple sub-controllers to partition a dynamic spectrum access problem into separate sub-problems, reducing the complexity of band selection. In \cite{he2022meta}, the authors deploy a Meta-HRL framework for resource allocation in vehicular networks to enable faster learning on newly discovered sub-tasks.

\subsection{Our Contributions}
In this paper, we propose novel methods to orchestrate parameters of \gls{urllc} services.  
Reference \cite{ganjalizadeh2023tii} developed a single-agent \gls{rl} method on a factory automation scenario to jointly optimize the \gls{dl} transmission power and \gls{harq} retransmission control for the communication service performance (availability and reliability) in \gls{urllc} services. Here, we address the same problem with a novel HRL framework that supports better performance but with a more flexible structure that enables the system to allocate multiple agents and execute different operations with multi-level policies in different timescales. Our solution substantially reduces the signaling requirements for training and inference. In particular, two of the agents are located at the \glspl{gnb}, which significantly reduces the data exchange between \glspl{gnb} and the centralized remote \gls{hrl} agent. This efficiency results in time and energy savings in decision-making, thus simplifying the adaptation of our framework to the complex network requirements of real-world wireless systems and 6G.

\noindent \textbf{Notations:} Normal font $x$ and $X$, bold font \boldsymbol{$x$}, and uppercase calligraphic font $\mathcal{X}$ denote scalars, vectors, and sets respectively. Besides, $[X]$ denotes the set $\{1, 2, \ldots, X\}$, and $|\mathcal{X}|$ is the cardinality of set $\mathcal{X}$.

\section{System Model and Performance Metrics}\label{sec:PR}
\subsection{System Model}

We consider a factory automation scenario where a set of $\calB {\coloneqq}[B]$ \glspl{gnb} are present, each serving a set of $\calU_b {\coloneqq} [U_b]$ industrial devices, where $\calU_b\subset\calU$, and $\calU{\coloneqq}[U]$ is the set of all devices. These devices are responsible for executing various functions that facilitate automated production. 
In this scenario, the communication system must be capable of delivering monitoring data to \glspl{gnb} and computed or emergency control commands to the actuators in a timely and reliable manner. 
For the channel model, we assume \gls{inf-dh} from 3GPP in~\cite{3gpp.38.901}. Nevertheless, our problem formulation and approach, described in Section~\ref{sec:hrl} and Section~\ref{sec:simulationresults}, are not limited to this channel model.
To enable \gls{urllc} efficiently, we consider the orchestration of a set of reliability enhancement features, such as the transmission power to industrial devices and the maximum number of diversity transmissions (i.e., transmitting multiple instances of a packet or its segments in space, time, and/or frequency).
In this paper, we focus on the orchestration of transmission power and \gls{harq} retransmissions in \gls{dl} direction. However, our framework can easily be extended for other reliability enhancement features and \gls{ul} transmissions.

From the network management perspective, in the hierarchical multi-tier architecture of cellular networks, we assume two levels of control for global and local optimizations. Although there is only one top-level controller, we assume a set of low-level controllers co-located with the \glspl{gnb}, managing the transmissions towards $\calU_b$ together.

\subsection{From Network to Communication Service Performance}\label{sec:nwstate}
The key element in characterizing service performance is defining service failures accurately. In~\cite{3gpp.22.104}, \gls{3gpp} defines survival time denoted as $T_{\mathrm{s}}$, as the duration for which an application can continue to function without receiving an expected packet. Therefore, a communication service failure occurs if no packets have been received by the reception entity for the duration of survival time. 

We can define the network layer state variable $Y_{b,u}{(t)}$ for industrial device $u$ associated to \gls{gnb} $b$ at time $t$, where $Y_{b,u}{(t)}$ is considered $0$ if the last packet fails to reach the communication interface within a specified delay bound due to decoding issues at lower layers, excessive retransmissions, or queuing delays, and $1$ otherwise. 
We can define the network state variable $Y_{b,u}{(t)}$ for the $u$th industrial device at time $t$, where $Y_{b,u}{(t)}$ is considered $0$ if the last packet fails to reach the communication interface within a specified delay bound due to decoding issues at lower layers, excessive retransmissions, or queuing delays, and $1$ if the packet is received successfully and timely.
Since sporadic packet losses\footnote{Here, packet loss refers to all packets that fail to reach their intended recipient within their delay bound.} within $T_{\mathrm{s}}$ do not impact the end-to-end service performance, the application layer state variable can be defined as~\cite{ganjalizadeh2023tii}:
\begin{equation}\label{eq:Z}
    Z_{b,u}{(t)}:=\max_{t-{T}_{\mathrm{s}}\leq\tau\leq t}Y_{b,u}{(\tau)}.
\end{equation}
The application state variable $Z_{b,u}$ enables us to define and formulate our two reliability \glspl{kpi}, communication service availability and communication service reliability.

\subsubsection*{Communication Service Availability} It refers to the ability of an end-to-end communication service to perform its intended function without failure at a given point in time and is commonly expressed as a probability or as a percentage of time that the system is operational within a specified time period~\cite{3gpp.22.261}.
Considering the failure definition and $Z_{b,u}{(t)}$ in \eqref{eq:Z}, the communication service availability, $\alpha_{b,u}$, is~\cite{ganjalizadeh2023tii}
\begin{equation}\label{eq:avail}
    \alpha_{b,u}\coloneqq\lim_{t\to\infty}\Pr(Z_{b,u}{\left(t\right)}=1)=\lim_{T\to\infty}\frac{1}{T}\int_{0}^{T}Z_{b,u}(t) dt. 
\end{equation}
However, the communication service availability for $u$th device can be approximated in a short time , $\Delta t_k$, via~\cite{ganjalizadeh2023tii}
\begin{equation}\label{eq:eavail}
        \overline{\alpha}_{b,u}(\Delta t_k)\coloneqq\frac{1}{\Delta t_k}\,\int^{t_k}_{t_k-\Delta t_k} Z_{b,u}{(t)} dt.
\end{equation}
\subsubsection*{Communication Service Reliability}  It refers to the ability of an end-to-end communication service to operate without failures over a specific period, given certain environmental and operational conditions~\cite{3gpp.22.104}. It can be expressed as the meantime that the service is operational, that is $Z_{b,u}{(t)} = 1$. Therefore, reliability $\rho_{b,u}$ is formulated as~\cite{ganjalizadeh2023tii}

\begin{equation}\label{eq:relia}
    \rho_{b,u}\coloneqq\lim_{T\to\infty}\frac{1}{F_{b,u}{\left(T\right)}}\int_{0}^{T}Z_{b,u}(t) dt,
\end{equation}
\noindent where $F_{b,u}{\left(T\right)}$ denotes the number of crossings from $Z_{b,u}{(t)} = 1$ to $Z_{b,u}{(t)} = 0$ within $[0,T]$.
Since communication service reliability's unit is time, we can alternatively approximate it via the crossing rate, $\psi_{b,u}$, representing the crossings of $Z_{b,u}(t)$ from $1$ to $0$ during $\Delta t_k$; defined as $\psi_{b,u} {\coloneqq} \lim\limits_{T\to\infty}F_{b,u}{(T)}/T$. Note that $\psi_{b,u}$ is inversely proportional to $\rho_{b,u}$ in \eqref{eq:relia}. Then, the crossing rate can be approximated by~\cite{hoyland2009system}
\begin{equation}\label{eq:erelia}
        \overline{\psi}_{b,u}(\Delta t_k)\coloneqq\frac{F_{b,u}(\Delta t_k)}{\Delta t_k}.
\end{equation}
In the following section, we introduce our \gls{hrl} solution based on the formulation of the \glspl{kpi}.
\section{Optimization with HRL Framework}\label{sec:hrl}
In this paper, the objective is to maximize the communication service availability, in \eqref{eq:avail}, and communication service reliability, in \eqref{eq:relia}, of a \gls{cps} by optimizing the configuration of transmission power levels and the number of retransmissions. Nevertheless, our framework can easily be extended for more decision variables.
We propose to solve this problem using a \gls{hrl} framework, where a high-level agent collaborates with low-level agents to manage the bi-level control of the communication system. The high-level agent is responsible for inter-agent coordination and, therefore, we assign the task of mitigating inter-cell interference globally to it by placing the transmission power under its control. 
Hence, we model the problem as a twin timescale \gls{mdp} and then apply the \gls{sac} algorithm, as presented in~\cite{haarnoja2018soft}, to solve it. Additionally, we use the branching technique described in~\cite{ganjalizadeh2023tii} to enhance the performance of the algorithm.
For simplicity, we assume that the timescales of the low-level agents are identical and denote it with $\Delta t_k$, for an iteration starting from $t_k$, and represent the high-level agent's timescale with $\Delta t_{\kappa}^{\mathrm{h}}\coloneqq t_{\kappa}-t_{\kappa{-}1}$, where $\Delta t_{\kappa}^{\mathrm{h}}=c\Delta t_k$, $\forall c, k,\kappa\in\mathbb{N}$.


\subsection{State Space}\label{sec:state}
The state represents the set of various measurements from the environment that affects the performance of our main \glspl{kpi}, namely communication service availability and reliability. The state of $u$th device associated with the $b$th low-level agent, $\bm{s}_{b,u}{(\Delta t_k)}$, is measured within $[t_k-\Delta t_k, t_k]$, and consists of various factors that can be classified as direct and indirect, based on their effects on the two \glspl{kpi}.
As mentioned, individual availability $\alpha_{b,u}$ and reliability $\rho_{b,u}$ are the functions of the probability of packet loss and average operational time. Therefore, we add packet loss rate and mean downtime of the network layer as the direct factors to the state space. Apart from these two variables, we also consider the various factors that are not included in the \gls{kpi} functions but importantly affect the communication quality indirectly, as \gls{sinr}, packet transmission delays, the status of the \gls{rlc} layer buffers, path gain, the number of \gls{harq} transmissions, and the number of used resource blocks.
To enable a more concrete description of the communication environment, we include mean, median values, $95$th, and $5$th percentile of \gls{sinr}, path gain, and \gls{rlc} buffer status. However, for the rest of the factors, we incorporate the mean of the samples in the state.
Thus, the state of $b$th low-level agent for its $k$th iteration is defined as $\bm{s}_{b}{(\Delta t_k)} {\coloneqq} \left\{\bm{s}_{b,u}{(\Delta t_k)}| \forall u\in \calU_b\right\}$.

As for the high-level agent, we define the global state consisting of all elements in $\bm{s}_{b,u}$, but  measured in a much larger time scale, $\Delta t_{\kappa}^{\mathrm{h}}$, and including all devices. Then, the high-level agent's state for $\kappa$th iteration can be defined as $\bm{s}{(\Delta t_{\kappa}^{\mathrm{h}})} {\coloneqq} \left\{\bm{s}_{b,u}{(\Delta t_{\kappa}^{\mathrm{h}})}| \forall u\in \calU_b, \forall b\in \calB\right\}$.

\subsection{Action and HRL Policy}\label{sec:action}

The action space consists of a series of decision parameters for the agents to interact with environments. Similar to reference~\cite{ganjalizadeh2023tii}, we consider the quantized transmission power levels and the number of retransmissions in action set for each device.
However, we decompose the joint action into a multi-level policy that enables the different \gls{hrl} agents to learn one or a combination of different network functions with different timescales based on network requirements. Specifically for our scenario, we set up a two-level policy that includes a global optimization of transmission powers (via a high-level agent), and local optimization of the maximum number of retransmissions (via several low-level agents). Hence, the high-level action in $\kappa$th iteration is defined as  $\boldsymbol{
a}_{\kappa}^{\mathrm{h}}\coloneqq(p^1_{\kappa},\ldots,p^{u}_{\kappa},\ldots,p^{U}_{\kappa})$, where $p^u_{\kappa} \in\{p_{\min},p_{1},p_{2},\ldots, p_{\max}\}$, and the action for $b$th low-level agent in $k$th iteration, $\boldsymbol{a}_{b, k}$, is a vector where each element represents the configured maximum numbers of retransmissions for devices in $\calU_b$. Compared to flat (or single-level) \gls{rl}, \gls{hrl} decomposes the action space into layers, resulting in higher scalability for handling complex orchestrations in cellular systems.
\begin{figure}[t]
\centering
\includegraphics[width=.87\columnwidth,keepaspectratio]{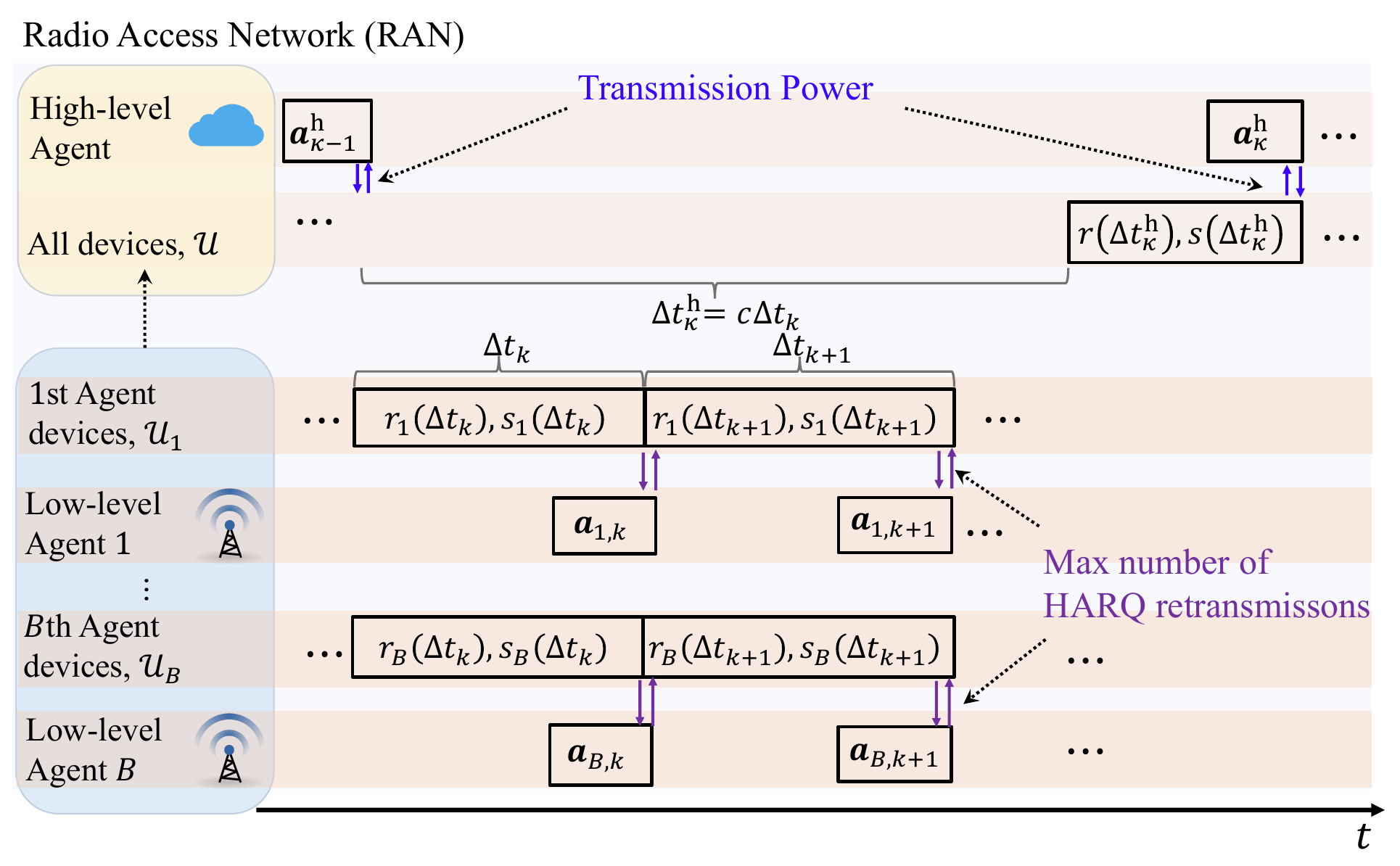}
\caption{Learning procedure of the two-level policy \gls{hrl} framework}
\label{fig:learningprocedure}
\vspace{-2mm}
\end{figure}

\subsection{Reward Functions}\label{sec:reward}
The \gls{rl} agents follow an explicit objective to maximize the sum of discounted rewards. 
Considering the estimations on communication service availability and crossing rate in \eqref{eq:eavail} and \eqref{eq:erelia}, respectively, we introduce two different reward functions. The first one, inspired by~\cite{ganjalizadeh2023tii}, targets the maximization of the average of reliability \glspl{kpi}, and is defined as
\begin{equation}\label{eq:rewAvg}
    r_{b}{(\Delta t_k)}\coloneqq \frac{1}{\omega U_b}\sum_{u=1}^{U_b} \left(\omega\overline{\alpha}_{b, u}(\Delta t_k)-(1{-}\omega)\overline{\psi}_{b, u}(\Delta t_k)\right),
\end{equation}
\noindent where $ 0{<}\omega{<}1$ decides the importance of $\overline{\alpha}_{b, u}(\Delta t_k)$ and $\overline{\psi}_{b, u}(\Delta t)$,  and its placement in the denominator helps to bound the reward function by $1$. Consequently, the high-level reward is defined as $r{(\Delta t_{\kappa}^{\mathrm{h}})}{\coloneqq}\frac{1}{B}\sum_{b=1}^{B}r_b{(\Delta t_{\kappa}^{\mathrm{h}})}$, where $r_b{(\cdot)}$ is derived as in \eqref{eq:rewAvg}, but within $(t_{\kappa{-}1},t_{\kappa}]$.
For the second function, we adopt a risk-sensitive reward that replaces the average value of users with the extremum within $\calU_b$, where agents can be more aggressive in exploration. Enabling agent $m$ to maximize the availability and minimize the crossing rate, the $k$th iteration reward is defined as
\begin{equation}\label{eq:rewRisk}
    r_b{(\Delta t_k)}\coloneqq \exp\left(\frac{\eta}{\omega}\left(r'_b{(\Delta t_k)} - \omega\right)\right),
\end{equation}
where
\begin{equation}\label{eq:rPrime}
   r'_b{(\Delta t_k)}\coloneqq\omega\min_{u \in \mathcal{U}_{b}}\!\!\left(\overline{\alpha}_{b, u}{(\Delta t_{k})}\right)-(1{-}\omega)\max_{u \in \mathcal{U}_{b}}\!\!\left(\overline{\psi}_{b,u}{(\Delta t_{k})}\right),
\end{equation}
and $\eta$ represents a fixed coefficient predefined to adjust the reward reduction with the application's sensitivity to reliability \glspl{kpi}.
The high-level reward, $r{(\Delta t_{\kappa}^{\mathrm{h}})}$, can then be calculated
as \eqref{eq:rewRisk}, but within $(t_{\kappa{-}1},t_{\kappa}]$ where the $\min$ and $\max$ functions in \eqref{eq:rPrime} are calculated for $\forall u\in\calU$.

\subsection{Learning Procedure}\label{sec:algo}
The learning procedure of our two-level policy \gls{hrl} framework is as shown in \figurename \ref{fig:learningprocedure}, where two network operations, global power control and maximum \gls{harq} retransmission number are decided by a high-level agent and low-level agents, parallelly following two timescales ($\Delta t_{k}$, $\Delta t^{\mathrm{h}}_{\kappa}$).
At every time step $\Delta t_{k}$, the low-level agents collect the states of the users they control, also the calculated rewards. Then the generated actions for \gls{harq} retransmission number are sent to the communication system. The high-level agent issues power values to all the devices every $c\Delta t_{k}$. Similar to the design in \cite{ganjalizadeh2021rl}, all the agents are model-free and can individually implement any \gls{rl} algorithms according to the task requirements. For the sake of fair comparison to the single-agent RL in~\cite{ganjalizadeh2023tii}, we deploy the same \gls{sac} algorithm \cite{haarnoja2018soft} to all the \gls{hrl} agents, and adopt the same branching technique that enables the continuous actions to describe our discrete actions in our factory automation scenario.



\section{Simulation Methodology and Result Analysis}\label{sec:simulationresults}

In this section, we present the simulation methodology and evaluate the performance of our \gls{hrl} framework.

\subsection{Simulation Configuration and Methodology}\label{sec:simulation}

\begin{figure}[t]
\centering
\includegraphics[width=.96\columnwidth,keepaspectratio]{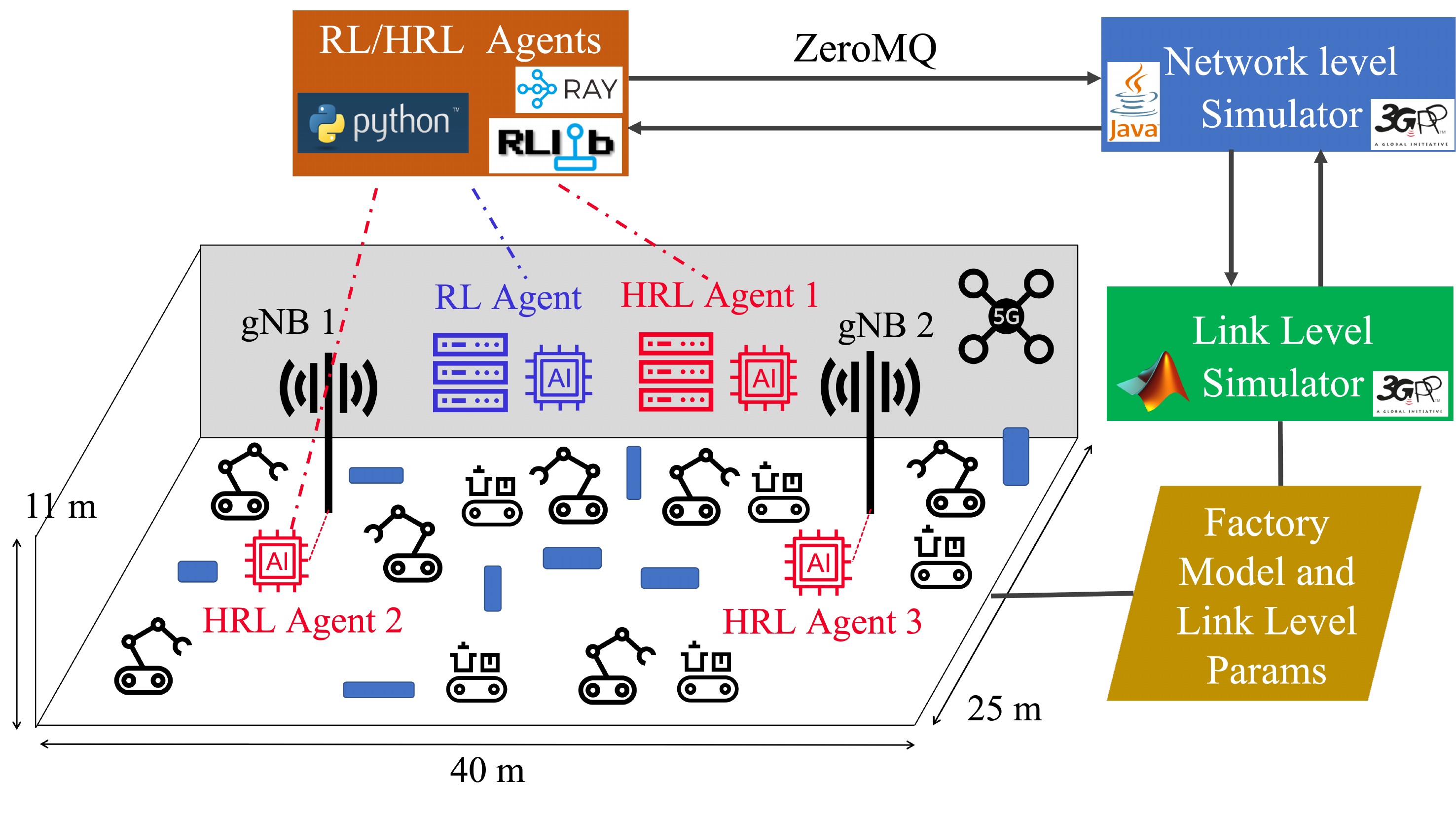}
\caption{Framework of the simulators with scenario setup}
\label{fig:simulator}
\vspace{-2mm}
\end{figure}

\figurename\,\ref{fig:simulator} shows our simulation architecture, which consists of a link-level Matlab simulator, a network-level Java simulator, and the \gls{rl} agents implemented using an open source library, RLlib. With path gain matrices, 3D channel data \cite{3gpp.38.901}, and nodes allocation provided by the link-level simulator, the network-level simulator is able to simulate the network with multiple \glspl{gnb} and users in \gls{phy}, \gls{mac}, and other higher layers of the \gls{5g}-NR network. As for the \gls{rl}/\gls{hrl} deployment, our network-level simulator supports interacting with external agents using the pipelines based on ZeroMQ protocol.

We considered a $40 \times 25 \times 11$\,m$^{3}$ factory with two $10$\,m height \glspl{gnb} providing communication service to $10$ industrial devices. Moreover, we assumed that the devices are in a high interference condition, and they move with the speed of $30$\,km/h, while staying in close proximity to
the original position. We considered periodic control traffic with periodicity $2$\,ms, and a delay bound $2.5$\,ms. On the transmitter side, packets are queued in the \gls{rlc} buffers and then sent via transport blocks based on the selected \gls{mcs}.
On the receiver side, the successfully decoded packets that were received after the delay bound were discarded by \gls{pdcp}. In the end, the reliability and availability are calculated in the application layer based on survival time, $T_{\mathrm{s}}$. The network and learning parameters of our simulations are presented in \tablename\,\ref{tab:netpara}. For more details on our simulation setup, you can refer to \cite[\S VI.A, \S VI.B]{deviceSelection}.
\begin{table}[t]
\centering
\caption{Simulation Parameters.}
\label{tab:netpara}
\scalebox{0.75}{\begin{tabular}{  l|l  }
	\hline
        \multicolumn{2}{c}{\textbf{Network Parameters}} \\
        \hline
	\textbf{Parameter}& \textbf{Value}\\
	\hline
	Deployment   & $2$ \glspl{gnb}, $1$ cell each\\
	\glspl{gnb} antenna height&   $8$\,m \\
	Devices' height &$1.5$\,m \\
	Carrier frequency & $2.6$\,GHz \\
	Bandwidth&   $20$\,MHz\,\\
	TTI length/Subcarrier spacing& $0.5$\,ms/$30$\,KHz  \\
	DL transmit power $(p_{min}/p_{max})$ &   $0.2$\,W/$0.5$\,W  \\
        Number of \gls{gnb}/Devices' antennas & $2/2$ \\
	\gls{dl}\,\gls{urllc}\,delay bound& $2.5$\,ms \\
	\gls{dl} \gls{urllc} Survival time ($T_{\mathrm{s}}$)& $5$\,ms \\
	Simulation time &  $10$\,s$/$Episode\\
	\hline
        \multicolumn{2}{c}{\textbf{Learning Parameters}} \\
        \hline
        \textbf{Parameter}& \textbf{Value}\\
	\hline
	Neural network hidden layers & $128 \times 128$  \\
        Activation function & ReLU  \\
        Loss function & MSE  \\
        Optimizer & mini-batch SGD\\
	Discount factor & 0.1 \\
        Batch size & 200 \\
	Learning rate & 0.0003 \\
        \gls{rl} step period ($\Delta t$) & $0.1$\,s \\
	\gls{hrl} step period low/high-level ($\Delta t_k$ / $\Delta t_{\kappa}^{\mathrm{h}}$) & $0.1$/$0.5$\,s, $\forall k,\kappa\in\mathbb{N}$ \\
	\hline
\end{tabular}}
 \vspace{-4.5mm}
\end{table}

\begin{figure*}[h!]
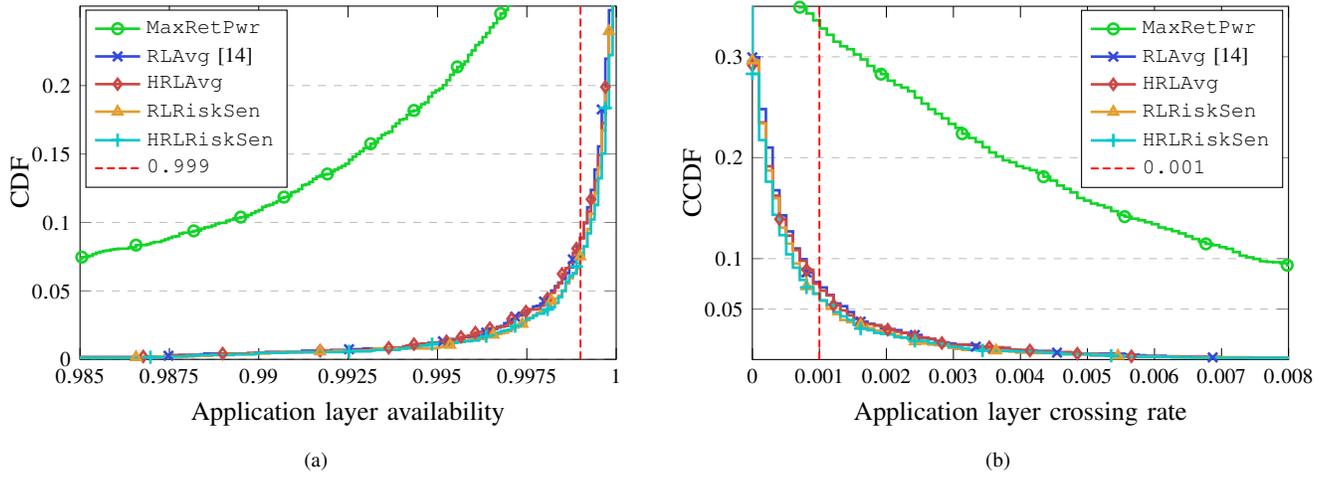

\centering
\begin{subfigure}[t]{0.5\linewidth}
\centering
    \input{Components/Figs/avail_avg_risk_compare_conference}
  \caption{}
  \label{fig:ava}
\end{subfigure}%
\hfill
\begin{subfigure}[t]{0.5\linewidth}
\centering
    \input{Components/Figs/ns_avg_risk_compare_conference}
    \caption{}
    \label{fig:ns}
\end{subfigure}
\caption{ Comparison of single-agent \gls{rl} solution and \gls{hrl} framework in terms of (a) communication service availability, and (b) crossing rate of \glspl{ue}, sampled from all the devices from the $10$\,s simulations.}
  \label{fig:ava_ns_twobs}
   \vspace{-2mm}
\end{figure*}

The goal of our \gls{hrl} solution is to find the optimal solution for power control and \gls{harq} retransmissions to maximize communication service availability and reliability. We considered the following baselines in our evaluations:
\begin{itemize}
    \item \texttt{MaxRetPwr}: Similiar to~\cite{ganjalizadeh2023tii}, all resource blocks are configured with $0.02$\,W, and the maximum number of transmissions is set to 2.
    
    \item \texttt{RLAvg}/\texttt{RLRiskSen}: There exist only one \gls{rl} agent (in a remote server) interacting with the two \glspl{gnb} (the blue agent in \figurename\,\ref{fig:simulator}). The step period is set to $0.1$\,s. The agent selects from the two possible levels of $0.008$\,W, and $0.02$\,W, set for all resource blocks allocated to a specific device, and the maximum number of transmissions (which can be either 1 or 2). While \texttt{RLAvg} used the average reward from \cite{ganjalizadeh2023tii} and presented in \eqref{eq:rewAvg}, \texttt{RLRiskSen} was implemented with the risk-sensitive reward in \eqref{eq:rewRisk}.
\end{itemize}
Furthermore, for the \gls{hrl} solution, we implemented one top-level agent and two low-level agents (the red agents in \figurename\,\ref{fig:simulator}). The former agent's objective was to optimize the \gls{dl} transmission powers globally, while the numbers of \gls{harq} retransmissions were optimized per \gls{gnb} by the latter agents. For the sake of fair comparison, the action space was set similar to \texttt{RLAvg} and \texttt{RLRiskSen}. Besides, we considered two setups as \texttt{HRLAvg}, where we incorporated the average reward in \eqref{eq:rewAvg}, and \texttt{HRLRiskSen}, where we incorporated the risk-sensitive reward, in \eqref{eq:rewRisk}.

\subsection{Result and Analysis}\label{sec:results}

\figurename\,\ref{fig:ava} and \figurename\,\ref{fig:ns} present the \gls{cdf} of the communication service availability and the \gls{ccdf} of the device crossing rate, determined by \eqref{eq:eavail} and \eqref{eq:erelia}, respectively. 
In these figures, each data point represents the device availability (in \figurename\,\ref{fig:ava}) or crossing rate (in \figurename\,\ref{fig:ns}) in one simulation round. 
Assuming an availability requirement of $0.999$, \figurename\,\ref{fig:ava} shows that both \texttt{HRLRiskSen} and \texttt{RLRiskSen} achieve a similar violation probability of $0.083$, while \texttt{MaxRetPwr} can only reach violation probability of $0.35$.
Similarly, assuming a crossing rate requirement of $0.001$, \figurename\,\ref{fig:ns} indicates that \texttt{HRLRiskSen} and \texttt{RLRiskSen} can obtain violation probability of $0.058$, closely followed by \texttt{HRLAvg}, and they all significantly outperform \texttt{MaxRetPwr}. Furthermore, our \gls{hrl} framework shows a lower violation probability for availability $>$ 0.999 and crossing rate $<$ 0.001, outperforming the \gls{rl}.

It is surprising that our \gls{hrl} framework achieves better performance than the single-agent \gls{rl} method, which learns the states from all the devices within one agent. In comparison, the \gls{hrl} agents have three different actions, where two of them only learn partial states of the devices from one \gls{gnb}. Since we run both \gls{rl} and \gls{hrl} training for a fixed number of iterations, such improvement can be contributed by the reduction in the action space (as a result of action decomposition), leading to faster convergence.

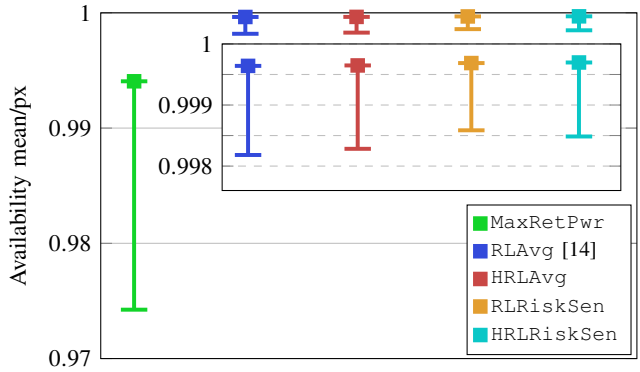
\begin{figure}[t]
    \centering
%
%
\definecolor{mycolor1}{rgb}{0.07059,0.83137,0.16078}%
\definecolor{mycolor2}{rgb}{0.03922,0.78039,0.78039}%
\definecolor{mycolor3}{rgb}{0.89020,0.61961,0.18824}%
\definecolor{mycolor4}{rgb}{0.20000,0.29020,0.85882}%
\definecolor{mycolor5}{rgb}{0.78824,0.27843,0.27059}%
\begin{tikzpicture}
\begin{pgfonlayer}{background}
\begin{axis}[%
width=72.0*0.99mm,
height=46.0mm,
at={(-0.1,0)},
scale only axis,
xmin=0.1,
xmax=2.51025865776113,
xtick={\empty},
ymin=0.97,
ymax=1,
ylabel style={font=\color{white!15!black}},
ylabel={Availability mean/px},
axis background/.style={fill=white},
xminorgrids,
ymajorgrids,
yminorgrids,
label style={font=\small},
tick label style={font=\small},
legend style={at={(0.99,0.01)}, anchor=south east, legend cell align=left, align=left, font=\footnotesize}
]
\addplot [color=mycolor1, line width=1.5pt, only marks, mark size=2pt, mark=square*, mark options={solid, mycolor1}]
 plot [error bars/.cd, y dir=both, y explicit, error bar style={line width=1.5pt}, error mark options={line width=1.5pt, mark size=5.0pt, rotate=90}]
 table[row sep=crcr, y error plus index=2, y error minus index=3]{%
0.25	0.994059917355371	0	0.0198174931129468\\
};
\addlegendentry{\texttt{MaxRetPwr}}

\addplot [color=mycolor4, line width=1.5pt, only marks, mark size=2pt, mark=square*, mark options={solid, mycolor4}]
 plot [error bars/.cd, y dir=both, y explicit, error bar style={line width=1.5pt}, error mark options={line width=1.5pt, mark size=5.0pt, rotate=90}]
 table[row sep=crcr, y error plus index=2, y error minus index=3]{%
0.75	0.999640852974188	0	0.00145903479236942\\
};
\addlegendentry{\texttt{RLAvg}~\cite{ganjalizadeh2023tii}}

\addplot [color=mycolor5, line width=1.5pt, only marks, mark size=2pt, mark=square*, mark options={solid, mycolor5}]
 plot [error bars/.cd, y dir=both, y explicit, error bar style={line width=1.5pt}, error mark options={line width=1.5pt, mark size=5.0pt, rotate=90}]
 table[row sep=crcr, y error plus index=2, y error minus index=3]{%
1.25	0.999648174870398	0	0.00136534658756937\\
};
\addlegendentry{\texttt{HRLAvg}}

\addplot [color=mycolor3, line width=1.5pt, only marks, mark size=2pt, mark=square*, mark options={solid, mycolor3}]
 plot [error bars/.cd, y dir=both, y explicit, error bar style={line width=1.5pt}, error mark options={line width=1.5pt, mark size=5.0pt, rotate=90}]
 table[row sep=crcr, y error plus index=2, y error minus index=3]{%
1.75	0.999686975575866	0	0.00110111699000726\\
};
\addlegendentry{\texttt{RLRiskSen}}

\addplot [color=mycolor2, line width=1.5pt, only marks, mark size=2pt, mark=square*, mark options={solid, mycolor2}]
 plot [error bars/.cd, y dir=both, y explicit, error bar style={line width=1.5pt}, error mark options={line width=1.5pt, mark size=5.0pt, rotate=90}]
 table[row sep=crcr, y error plus index=2, y error minus index=3]{%
2.25 0.999696542140987   0  0.001211693656138\\
};
\addlegendentry{\texttt{HRLRiskSen}}

\coordinate (inset) at (axis description cs:0.97,0.91);
\end{axis}
\end{pgfonlayer}

\begin{axis}[%
at={(inset)},
anchor=north east,
axis background/.style={fill=white!10},
width=53mm,
height=19.5mm,
scale only axis,
xmin=0.5,
xmax=2.5,
xtick={\empty},
ymin=0.9976,
label style={font=\small},
tick label style={font=\small},
ymax=1,
ytick = {0.998, 0.9985, 0.999,0.9995, 1},
yticklabels = {$0.998$, $ $, $0.999$, $ $, $1$},
 grid style={dashed},
 ymajorgrids,
]

\addplot [color=mycolor4, line width=1.5pt, only marks, mark size=2pt, mark=square*, mark options={solid, mycolor4}, forget plot]
 plot [error bars/.cd, y dir=both, y explicit, error bar style={line width=1.5pt}, error mark options={line width=1.5pt, mark size=5.0pt, rotate=90}]
 table[row sep=crcr, y error plus index=2, y error minus index=3]{%
0.63	0.999640852974188	0	0.00145903479236942\\
};

\addplot [color=mycolor5, line width=1.5pt, only marks, mark size=2pt, mark=square*, mark options={solid, mycolor5}, forget plot]
 plot [error bars/.cd, y dir=both, y explicit, error bar style={line width=1.5pt}, error mark options={line width=1.5pt, mark size=5.0pt, rotate=90}]
 table[row sep=crcr, y error plus index=2, y error minus index=3]{%
1.18	0.999648174870398	0	0.00136534658756937\\
};

\addplot [color=mycolor3, line width=1.5pt, only marks, mark size=2pt, mark=square*, mark options={solid, mycolor3}, forget plot]
 plot [error bars/.cd, y dir=both, y explicit, error bar style={line width=1.5pt}, error mark options={line width=1.5pt, mark size=5.0pt, rotate=90}]
 table[row sep=crcr, y error plus index=2, y error minus index=3]{%
1.75	0.999686975575866	0	0.00110111699000726\\
};

\addplot [color=mycolor2, line width=1.5pt, only marks, mark size=2pt, mark=square*, mark options={solid, mycolor2}, forget plot]
 plot [error bars/.cd, y dir=both, y explicit, error bar style={line width=1.5pt}, error mark options={line width=1.5pt, mark size=5.0pt, rotate=90}]
 table[row sep=crcr, y error plus index=2, y error minus index=3]{%
2.29	0.999696542140987	0	0.001211693656138\\
};

\end{axis}

\end{tikzpicture}%
    \caption{Mean (square) and 5th percentile (line) availability for the simulations}
    \label{fig:bar_twobs}
    \vspace{-2mm}
\end{figure}

The error bar plot in \figurename\,\ref{fig:bar_twobs} demonstrates the mean (shown by square) and 5th percentile (shown by line) device availability. The average and 5th percentile availability of single-agent \gls{rl} and \gls{hrl} in both rewards achieve similar performance that is much higher than \texttt{MaxRetPwr}. Although the baseline simulation adopts the maximum power and 2 \gls{harq} retransmissions, its poor performance indicates that the lack of freedom in operations could reduce interference management capability. Same as in \figurename\,\ref{fig:ava_ns_twobs}, the risk-sensitive reward shows improvement in performance than the average reward for the \gls{hrl} and \gls{rl}.

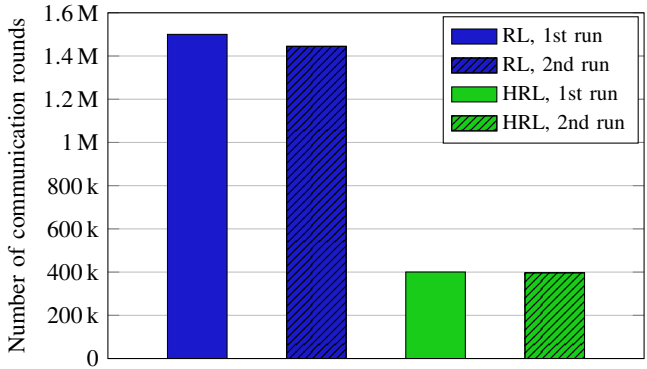
\begin{figure}[t]
    \centering
%
%
\definecolor{mycolor1}{rgb}{0.1,0.1,0.8}%
\definecolor{mycolor2}{rgb}{0.1,0.8,0.1}%
\begin{tikzpicture}

\begin{axis}[%
width=72.0*0.99mm,
height=46.0mm,
at={(1.011in,0.692in)},
scale only axis,
scaled y ticks = false,
xmin=-0.2,
xmax=1.6,
xtick={\empty},
ytick={0,200000,400000,600000,800000,1000000,1200000,1400000,1600000},
yticklabels={0, $200$\,k, $400$\,k, $600$\,k, $800$\,k, $1$\,M, $1.2$\,M, $1.4$\,M, $1.6$\,M},
xlabel style={font=\color{white!15!black}},
ymin=0,
ymax=1600000,
ylabel style={font=\color{white!15!black}},
ylabel={Number of communication rounds},
axis background/.style={fill=white},
label style={font=\small},
tick label style={font=\small},
legend style={at={(0.99,0.99)}, anchor=north east, legend cell align=left, align=left},
xmajorgrids,
ymajorgrids
]
\addplot[ybar, bar width=0.2, fill=mycolor1, area legend] table[row sep=crcr] {%
0.1	1500096\\
};
\addlegendentry{\footnotesize RL, 1st run}

\addplot[ybar, bar width=0.2, fill=mycolor1, area legend,postaction={pattern=north east lines}] table[row sep=crcr] {%
0.5	1444720\\
};
\addlegendentry{\footnotesize RL, 2nd run}

\addplot[ybar, bar width=0.2, fill=mycolor2, area legend] table[row sep=crcr] {%
0.9	400400\\
};
\addlegendentry{\footnotesize HRL, 1st run}

\addplot[ybar, bar width=0.2, fill=mycolor2, area legend,postaction={pattern=north east lines}] table[row sep=crcr] {%
1.3	396784\\
};
\addlegendentry{\footnotesize  HRL, 2nd run}

\end{axis}
\end{tikzpicture}%
    \caption{Number of signal exchanges in the training process.}
    \label{fig:sig_num}
    \vspace{-1mm}
\end{figure}
 
\figurename\,\ref{fig:sig_num} presents the number of signal exchanges (communication rounds) the learning frameworks require to converge (i.e., all reward values tend to stabilize and converge, and they no longer increase with further exploration). One signal exchange in \figurename\,\ref{fig:sig_num}  represents a signal transmission between the remote agent and one of the \glspl{gnb}. According to the allocation of agents in \figurename\,\ref{fig:simulator}, there are two learning agents assembled locally with the \glspl{gnb}, the transmission overhead of which can be ignored compared to the remote server. In the case of the single-agent \gls{rl} solution, the agent transmits one message of action to each \gls{gnb} and receives a reply message with states and rewards from them. Therefore, there are four signal transmissions at every step. Similarly, for \gls{hrl} framework, there are four signal transmissions at every high-level step. We performed two \gls{rl} simulations and two \gls{hrl} simulations and logged the total number of exchanged signals till convergence. 
As \figurename\,\ref{fig:sig_num} confirms, compared to \gls{hrl}, \gls{rl} simulations required over triple the number of communication rounds to converge.
Such reduction in signal exchanges can significantly improve the energy efficiency of large-scale communication systems with many remote devices and simultaneous operations.
Beyond signal exchanges, decomposing the \gls{rl} action allows the \gls{hrl}'s average learning time per iteration to be 33\% less than that of the single-agent \gls{rl} (i.e., $6.288$\,ms vs $8.386$\,ms).
This translates into massive gains in terms of latency and energy saving to run the training, especially in dynamic environments where we may need to retrain the models regularly.



\section{Conclusions}\label{sec:conculsions}

In this paper, we propose a \gls{hrl} framework and implement that into a simulated factory automation model to optimize the operations of power control and \gls{harq} retransmissions in \gls{5g} communication, aiming to achieve the optimal availability and reliability according to the standard of \gls{urllc}. We design and compare five simulations that separately deploy the single-agent \gls{rl} strategy, our \gls{hrl} framework with average and risk-sensitive reward functions, and one with fixed operations as the baseline. Our \gls{hrl} framework achieves better performance on availability and reliability to the ideal single-agent \gls{rl} solution and significantly outperforms the baseline simulation. Besides, our \gls{hrl} framework enables better flexibility that allows the operations to be executed in different timescales. Furthermore, due to the flexible allocation of agents, the \gls{hrl} solution can save signal consumption by at least
triple less than that of the \gls{rl} 
in the scenario of our factory model.

\end{NoHyper}
\ifCLASSOPTIONcaptionsoff
\newpage
\fi

%
%
\Urlmuskip=0mu plus 1mu\relax
\bibliographystyle{IEEEtran}


\bibliography{Components/Metafiles/IEEEabrv,Components/Metafiles/ref}
\clearpage
\end{document}